\documentclass[amsmath,amssymb,nofootinbib,notitlepage,superscriptaddress]{revtex4-1}

\usepackage{aas_macros}
\usepackage[colorlinks=true]{hyperref}

\let\originalleft\left
\let\originalright\right
\renewcommand{\left}{\mathopen{}\mathclose\bgroup\originalleft}
\renewcommand{\right}{\aftergroup\egroup\originalright}
\mathcode`\*="8000
{\catcode`\*=\active\gdef*{\mathclose{}\,\mathopen{}}}

\newcommand{\br}[1]{\left[#1\right]}

\newcommand{\pa}[1]{\left(#1\right)}

\newcommand{\ed}{\mathop{}\!\mathrm{d}}
\newcommand{\pd}{\mathop{}\!\partial}

\begin{document}

\title{\LARGE{Why there is no Love in black holes}}

\author{Alex Lupsasca}
\affiliation{Vanderbilt University}

\begin{abstract}
This paper presents a new conformal  symmetry of stationary, axisymmetric Kerr perturbations.
This symmetry is exact but non-geometric (or ``hidden''), and each of its generators has an associated infinite family of eigenstate solutions.
Tidal perturbations of a black hole form an irreducible highest-weight representation of this conformal group, while the tidal response fields live in a different such representation.
This implies that black holes have no tidal deformability, or vanishing Love numbers. 
\end{abstract}

\maketitle

\section{Introduction}

The spacetime geometry of a Kerr black hole is completely determined by its mass $M$ and angular momentum $J$.
In terms of Boyer-Lindquist coordinates $(t,r,x,\phi)$ with $x=\cos{\theta}$ as the polar angle, the Kerr line element is
\begin{subequations}
\label{eq:Kerr}
\begin{gather}
    ds^2=-\frac{\Delta}{\Sigma}\br{\ed t-a\pa{1-x^2}\ed\phi}^2+\frac{1-x^2}{\Sigma}\br{\pa{r^2+a^2}\ed\phi-a\ed t}^2+\frac{\Sigma}{\Delta}\ed r^2+\frac{\Sigma}{1-x^2}\ed x^2,\\
    \Delta(r)=r^2-2Mr+a^2,\qquad
    \Sigma(r,x)=r^2+a^2x^2,
\end{gather}
\end{subequations}
where $a=J/M$ is the specific angular momentum of the black hole.
Teukolsky famously showed that massless spin-$s$ perturbations of the Kerr black hole are described by two scalar fields $\Psi^{(\pm s)}$ that obey the same decoupled equation.
For vacuum perturbations, this is the ``Teukolsky master equation'' $\mathcal{T}^{(s)}\Psi^{(\pm s)}(t,r,x,\phi)=0$ with (see Eq.~(4.7) of \cite{Teukolsky1973})
\begin{align}
    \label{eq:TME}
    \mathcal{T}^{(s)}&=\br{\frac{\pa{r^2+a^2}^2}{\Delta}-a^2\pa{1-x^2}}\pd_t^2+\frac{4aMr}{\Delta}\pd_t\pd_\phi+\pa{\frac{a^2}{\Delta}-\frac{1}{1-x^2}}\pd^2_\phi\nonumber\\
    &\quad-2s\br{\frac{M\pa{r^2-a^2}}{\Delta}-\pa{r+iax}}\pd_t-2s\br{\frac{a(r-M)}{\Delta}+\frac{ix}{1-x^2}}\pd_\phi\nonumber\\
    &\quad-\Delta^{-s}\pd_r\pa{\Delta^{s+1}\pd_r}-\pd_x\br{\pa{1-x^2}\pd_x}+\pa{\frac{x^2s^2}{1-x^2}-s}.
\end{align}
When $s=0$, this equation reduces to the simple wave equation for a massless scalar field $\psi\equiv\Psi^{(0)}$,
\begin{align}
	\label{eq:KerrLaplacian}
	\nabla^2\psi(t,r,x,\phi)=0.
\end{align}
This paper presents four new results:
\begin{enumerate}
	\item Sec.~\ref{sec:Symmetry} shows that perturbations of a Kerr black hole exhibit an emergent conformal symmetry in the stationary and axisymmetric limit.
	More precisely, in that regime, the space of solutions to Eq.~\eqref{eq:TME} admits three continuous symmetries that map solutions to solutions, and that together generate the global conformal group $\mathsf{SL}(2,\mathbb{R})$.
	The stationary, axisymmetric equation itself can be exactly expressed as the Casimir of this symmetry group, which is ``hidden'' in the sense that it cannot be geometrically realized via the action of vector fields or isometries.
	\item Sec.~\ref{sec:Separability} shows that each of the three $\mathsf{SL}(2,\mathbb{R})$ generators is associated with an infinite family of exact solutions that diagonalize it.
	These solutions are not separable in the usual coordinates, but each symmetry generator is associated with a different coordinate system in which the solutions that diagonalize it do become ``$R$-separable''.
	\item Sec.~\ref{sec:Origin} shows that in these coordinates, the stationary and axisymmetric version of Eq.~\eqref{eq:KerrLaplacian} reduces to the usual Laplace equation $\nabla^2\psi(\rho,z)=0$ for an axisymmetric massless scalar in 3D flat space with cylindrical coordinates.
	In other words, stationary, axisymmetric fields on Kerr behave identically to their flat-space analogues, so that the presence of the black hole can be ``removed'' by a simple coordinate transformation!
	One of the symmetries (with generator $H_+$) generalizes translations $\pd_z$ along the flat-space cylindrical axis, but the others are novel.
	\item Sec.~\ref{sec:NoLove} uses this symmetry to show that black holes have no tidal deformability (or ``vanishing Love numbers'').	
	The global conformal group contains ``dilatations'' with generator $H_0$.
	The associated solutions $\psi_h$ are eigenstates $H_0\psi_h=(h+\frac{1}{2})\psi_h$ labeled by a conformal weight $h$.
	These modes all fall into two irreducible, highest-weight representations of $\mathsf{SL}(2,\mathbb{R})$: one containing the positive-weight solutions, and the other the negative-weight ones.
	Physically, the $h>0$ modes correspond to exterior multipole moments, and hence to tidal perturbations.
	By contrast, the $h<0$ modes are interior multipole moments corresponding to tidal responses.
	Since applied tidal fields and induced tidal responses live in distinct representations, the Love numbers must vanish in axisymmetry.
\end{enumerate}
\textit{Note:} As this work was nearing completion, Penna independently released similar results for the scalar case $s=0$ \cite{Penna2025}.

\section{New symmetries of Kerr perturbations}
\label{sec:Symmetry}

The Kerr geometry \eqref{eq:Kerr} is stationary and axisymmetric.
Accordingly, it admits two Killing vector fields $\partial_t$ and $\partial_\phi$.
Perturbations of the Kerr background that are also stationary and axisymmetric can be described by Teukolsky scalars $\Psi^{(\pm s)}$ that are invariant under these isometries, meaning that they also satisfy the two additional conditions
\begin{align}
	\label{eq:Stataxi}
	\pd_t\Psi^{(s)}(t,r,x,\phi)=0,\qquad
	\pd_\phi\Psi^{(s)}(t,r,x,\phi)=0.
\end{align}
In Boyer-Lindquist coordinates, this simply means that the fields can be assumed to have no dependence on the time and azimuthal coordinates $(t,\phi)$, reducing the Teukolsky master equation \eqref{eq:TME} to the simpler form
\begin{align}
    \label{eq:StataxiTME}
    \br{\Delta^{-s}\pd_r\pa{\Delta^{s+1}\pd_r}+\pd_x\br{\pa{1-x^2}\pd_x}-\pa{\frac{x^2s^2}{1-x^2}-s}}\Psi^{(s)}(r,x)=0.
\end{align}
In particular, stationary and axisymmetric solutions of the massless wave equation \eqref{eq:KerrLaplacian} obey the even simpler equation
\begin{align}
    \label{eq:StataxiLaplacian}
    \pd_r\br{\Delta\pd_r\psi(r,x)}+\pd_x\br{\pa{1-x^2}\pd_x\psi(r,x)}=0.
\end{align}

In Sec.~\ref{sec:Origin}, Lie's ``reduction of order'' algorithm is used to derive the complete symmetry group of these equations.
Surprisingly, it turns out that they admit an exact conformal symmetry generated by the linear differential operators%
\begin{subequations}
\label{eq:Generators}
\begin{align}
	H_+&=\frac{x\Delta\pd_r+(r-M)\pa{1-x^2}\pd_x+s(r-M)x}{(r-M)^2-\pa{M^2-a^2}x^2},\\
	H_0&=\frac{(r-M)\Delta\pd_r+\pa{M^2-a^2}x\pa{1-x^2}\pd_x+s\pa{M^2-a^2}x^2}{(r-M)^2-\pa{M^2-a^2}x^2}+\pa{s+\frac{1}{2}},\\
	H_-&=\frac{\pa{M^2-a^2}x\Delta\pd_r-(r-M)\pa{1-x^2}\br{\Delta-\pa{M^2-a^2}x^2}\pd_x+s\pa{M^2-a^2}(r-M)x}{(r-M)^2-\pa{M^2-a^2}x^2}\nonumber\\
	&\quad+x\Delta\pd_r+(s+1)(r-M)x.
\end{align}
\end{subequations}
Indeed, one can explicitly check that each of these three operators maps any solution of Eq.~\eqref{eq:StataxiTME} into another solution.
These three differential operators are regular everywhere outside the horizon, and they obey the commutation relations
\begin{align}
	\label{eq:SL2R}
	\br{H_0,H_\pm}=\mp H_\pm,\qquad
	\br{H_+,H_-}=2H_0,
\end{align}
which define the Lie algebra of the global conformal group $\mathsf{SL}(2,\mathbb{R})$.
Its Casimir $\mathcal{C}$ is the quadratic differential operator that commutes with all the generators of the symmetry group,
\begin{align}
	\label{eq:CasimirSL2R}
	\mathcal{C}=-H_0^2+\frac{H_+H_-+H_-H_+}{2}.
\end{align}
This Casimir is equivalent to the stationary, axisymmetric Teukolsky master equation \eqref{eq:StataxiTME}, which can be rewritten as
\begin{align}
	\label{eq:Casimir}
	\pa{\mathcal{C}-\frac{1}{4}+s^2}\Psi^{(s)}(r,x)=0.
\end{align}
This is the first main result of this paper: stationary and axisymmetric perturbations of a Kerr black hole are governed by a new global conformal symmetry group generated by the operators \eqref{eq:Generators}.
Moreover, this $\mathsf{SL}(2,\mathbb{R})$ symmetry is exact, in the sense that the entire equation \eqref{eq:StataxiTME} is equivalent to the Casimir of the group [Eq.~\eqref{eq:Casimir}]; in particular, this implies that the action of the group is solution-generating, mapping physical solutions into other physical solutions.

This symmetry may be called ``hidden'' because it is not geometrically realized as an isometry of the Kerr spacetime.
Stationarity and axisymmetry correspond to Killing vector fields $\partial_t$ and $\partial_\phi$ that act on physical fields via the Lie derivative $\mathcal{L}_V$: in Eq.~\eqref{eq:Stataxi}, one could have written $\mathcal{L}_{\pd_t}\Psi^{(s)}=\mathcal{L}_{\pd_\phi}\Psi^{(s)}=0$.
By contrast, none of the generators $H_i$ in Eq.~\eqref{eq:Generators} has an action that is equivalent to the Lie derivative $\mathcal{L}_{h_i}$ along some vector field $h_i$, essentially because of the multiplicative terms that involve no differentiation.
The sole exception arises in the scalar case $s=0$, for which the multiplicative piece in $H_+$ vanishes, allowing one to define a vector field $h_+$ such that $H_+\psi=\mathcal{L}_{h_+}\psi$.
This symmetry can be geometrically interpreted as a generalized axial translation, as discussed in Sec.~\ref{sec:Origin} below.

\section{New separable solutions}
\label{sec:Separability}

The second main point of this paper is to provide analytic expressions for the infinite families of exact solutions to Eq.~\eqref{eq:StataxiTME} that diagonalize each of the $\mathsf{SL}(2,\mathbb{R})$ generators \eqref{eq:Generators}.
Each one of these families of solutions is not separable with respect to the usual poloidal coordinates $(r,x=\cos{\theta})$, but is instead separable in a different coordinate system.
This is consistent with the statement that every second-order differential operator $\mathcal{D}$ that commutes with Eq.~\eqref{eq:StataxiTME} has an associated coordinate system in which the equation ``$R$-separates'' (that is, separates up to an overall factor) \cite{Kalnins2018}.

A well-known example in the context of black hole perturbation theory is provided by the Carter tensor \cite{Carter1968,Walker1970}
\begin{align}
	\label{eq:KillingTensor}
	K_{\mu\nu}\ed X^\mu\ed X^\nu=\frac{r^2\pa{1-x^2}}{\Sigma}\br{a\ed t-\pa{r^2+a^2}\ed\phi}^2+\frac{a^2x^2\Delta}{\Sigma}\br{\ed t-a\pa{1-x^2}\ed\phi}^2-\Sigma\br{\frac{a^2x^2}{\Delta}\ed r^2+\frac{r^2}{1-x^2}\ed x^2},
\end{align}
the unique (with $g_{\mu\nu}$) irreducible rank-2 Killing tensor of the Kerr metric \eqref{eq:Kerr}.
By the Killing equation $\nabla_{(\sigma}K_{\mu\nu)}=0$, the associated second-order differential operator $\mathcal{K}\equiv\nabla_\mu K^{\mu\nu}\nabla_\nu$ commutes with the Laplacian operator $\nabla^2\equiv\nabla_\mu g^{\mu\nu}\nabla_\nu$:
\begin{align}
	\br{\mathcal{K},\nabla^2}=0.
\end{align}
As reviewed below Eq.~(3.76) of \cite{Frolov2017}, this can be viewed as the reason why the Kerr Laplacian \eqref{eq:KerrLaplacian} is separable in $(r,x)$.
In fact, this separability extends to the full Teukolsky master equation \eqref{eq:TME}, which admits a mode decomposition using Heun functions, but the connection to symmetries is not fully understood (see, e.g., \cite{Berens2024} and references therein).
For simplicity, this section restricts attention to the scalar case $s=0$, with the case of general spin deferred to Sec.~\ref{subsec:HigherSpinModes}.

\subsection{Spin-0 perturbations: usual modes}

Scalar, stationary, and axisymmetric Kerr perturbations obey Eq.~\eqref{eq:StataxiLaplacian}.
The separable solutions take the simple form
\begin{align}
	\label{eq:StandardModes}
	\psi_\ell(r,x)=\br{a_\ell P_\ell\pa{\frac{r-M}{\sqrt{M^2-a^2}}}+b_\ell Q_\ell\pa{\frac{r-M}{\sqrt{M^2-a^2}}}}P_\ell(x),
\end{align}
where the $P_\ell(u)$ are just the Legendre polynomials while the $Q_\ell(u)$ are the Legendre functions of the second kind.
Regularity at the poles requires that $\ell\in\mathbb{N}$ be an integer, which may be interpreted as the multipole moment number.
The $P_\ell$ modes are regular at the horizon $u=1$ (but not as $u\to\infty$), whereas the $Q_\ell$ modes are regular at infinity (but not at the horizon). 
The Kerr multipole expansion will be revisited in Sec.~\ref{sec:NoLove} below.

The mode solutions \eqref{eq:StandardModes} diagonalize the second-order differential operator $\mathcal{K}$,
\begin{align}
	\mathcal{K}\psi_\ell(r,x)=-\ell(\ell+1)\psi_\ell(r,x),
\end{align}
and one may interpret the eigenvalue $\ell(\ell+1)$ as a conserved ``total angular momentum'' in Kerr.
For a Schwarzschild black hole, which has spherical symmetry, this statement is exactly true: the metric \eqref{eq:Kerr} with $a=0$ has an $\mathsf{SO}(3)$ isometry group generated by the Killing vectors $J_x=\sin{\phi}\pd_\theta+\cot{\theta}\cos{\phi}\pd_\phi$, $J_y=\sin{\phi}\pd_\theta-\cot{\theta}\sin{\phi}\pd_\phi$, and $J_z=\pd_\phi$, which obey the usual commutators $\br{J_i,J_j}=\epsilon_{ijk}J_k$; in this special case, the Carter tensor \eqref{eq:KillingTensor} reduces to
\begin{align}
	\label{eq:Reducibility}
	K^{\mu\nu}=J_x^\mu J_x^\nu+J_y^\mu J_y^\nu+J_z^\mu J_z^\nu,
\end{align}
which is the Casimir of $\mathsf{SO}(3)$.
As such, in Schwarzschild, the differential operator $\mathcal{K}$ corresponds to the conserved angular momentum $|\vec{J}|^2$.
By contrast, Kerr has no spherical symmetry and hence no analogue of $\vec{J}$, but nevertheless the Carter tensor still provides an analogue of $|\vec{J}|^2$.
However, since $K_{\mu\nu}$ is irreducible for $a\neq0$, this higher-rank symmetry is not associated with any vector fields and cannot be thought of in terms of isometries of the spacetime.\footnote{A notable exception arises in the extremal Kerr case $a=M$, in which a certain near-horizon limit yields a spacetime geometry that is a vacuum solution of the Einstein equations in its own right; this so-called NHEK geometry has an $\mathsf{SL}(2,\mathbb{R})\times\mathsf{U}(1)$ isometry group \cite{Bardeen1999} and in this limiting geometry, the Carter tensor once again becomes reducible to the Casimir of the isometry group (see, e.g., \cite{Lupsasca2020}).}

\subsection{Spin-0 perturbations: new symmetry eigenmodes}

By definition, each of the generators \eqref{eq:Generators} commutes with the $\mathsf{SL}(2,\mathbb{R})$ Casimir $\mathcal{C}$ defined in Eq.~\eqref{eq:CasimirSL2R}.
Therefore, the second-order differential operators $H_i^2$ must also commute with $\mathcal{C}$, or equivalently [via Eq.~\eqref{eq:Casimir}] with Eq.~\eqref{eq:StataxiLaplacian}.
As such, there ought to exist three new coordinate systems in which this equation is $R$-separable.
It is possible to derive these coordinates by solving for the modes that diagonalize each of the $H_i$ in turn.
In Sec.~\ref{sec:Origin}, these coordinate systems will help elucidate the origin of the hidden conformal symmetry, while the eigenmodes of $H_0$ will be crucial in Sec.~\ref{sec:NoLove}.

\subsubsection{Dilatations}

To diagonalize the ``dilatations'' generated by $H_0$, one searches for solutions to Eq.~\eqref{eq:StataxiLaplacian} that are also $H_0$-eigenstates:
\begin{align}
	H_0\psi_h(r,x)=\pa{h+\frac{1}{2}}\psi_h(r,x).
\end{align}
This leads to a family of solutions parameterized by a conformal weight $h$,
\begin{align}
	\label{eq:EigenModesH0}
	\psi_h(r,x)=\br{\sqrt{\pa{1-x^2}\Delta+(r-M)^2x^2}}^hP_h\pa{\frac{(r-M)x}{\sqrt{\pa{1-x^2}\Delta+(r-M)^2x^2}}}.
\end{align}
These modes are not separable in $(r,x)$, but they are separable in the coordinates $(R,X)$ associated with $\mathcal{D}=H_0^2$,
\begin{align}
	\label{eq:SphericalKerr}
	R=\sqrt{\pa{1-x^2}\Delta+(r-M)^2x^2},\qquad
	X=\frac{(r-M)x}{\sqrt{\pa{1-x^2}\Delta+(r-M)^2x^2}},
\end{align}
in terms of which they take the very simple and separable form
\begin{align}
	\psi_h(R,X)=R^hP_h\pa{X}.
\end{align}
$R$ and $X$ are themselves eigenmodes of $H_0$ with conformal weights $h=1$ and $h=0$, respectively.
Crucially, the mode solutions \eqref{eq:EigenModesH0} are regular everywhere outside the horizon; this time, replacing $P_h$ by $Q_h$ leads to irregular solutions.

\subsubsection{Translations}

Next, by diagonalizing the ``translation'' generator $H_+$,
\begin{align}
	H_+\psi_+(r,x)=-\lambda\psi_+(r,x),
\end{align}
one is led to a family of regular solutions of Eq.~\eqref{eq:StataxiLaplacian} that are expressible in terms of the Bessel function $J_0(z)$,
\begin{align}
	\psi_+(r,x)=J_0\pa{\lambda\sqrt{\pa{1-x^2}\Delta}}e^{-\lambda(r-M)x}.
\end{align}
These solutions are not separable in $(r,x)$, but they are separable in the coordinates $(\Pi,Z)$ associated with $\mathcal{D}=H_+^2$,
\begin{align}
	\label{eq:CartesianKerr}
	\Pi=\sqrt{\pa{1-x^2}\Delta}
	=R\sqrt{1-X^2},\qquad
	Z=(r-M)x
	=RX
\end{align}
in terms of which they take the simpler and separable form
\begin{align}
	\psi_+(\Pi,Z)=J_0(\lambda\Pi)e^{-\lambda Z}.
\end{align}

\subsubsection{Special conformal transformations}

Finally, by diagonalizing the generator $H_-$ of ``special conformal transformations'' (a name to be justified in Sec.~\ref{sec:Origin}),
\begin{align}
	H_-\psi_-(r,x)=\lambda\psi_-(r,x),
\end{align}
one is led to another family of regular solutions of Eq.~\eqref{eq:StataxiLaplacian} that are expressible in terms of the Bessel function $J_0(z)$,
\begin{align}
	\psi_-(r,x)=\frac{1}{\sqrt{\pa{1-x^2}\Delta+(r-M)^2x^2}}J_0\pa{\frac{\lambda\sqrt{\pa{1-x^2}\Delta}}{{\pa{1-x^2}\Delta+(r-M)^2x^2}}}e^{-\frac{\lambda(r-M)x}{\pa{1-x^2}\Delta+(r-M)^2x^2}}.
\end{align}
These solutions are not separable in $(r,x)$, but they are separable in the coordinates $(\tilde{\Pi},\tilde{Z})$ associated with $\mathcal{D}=H_-^2$,
\begin{align}
	\label{eq:InversionCoordinates}
	\tilde{\Pi}=\frac{\sqrt{\pa{1-x^2}\Delta}}{{\pa{1-x^2}\Delta+(r-M)^2x^2}}
	=\frac{\Pi}{\Pi^2+Z^2}
	=\frac{\Pi}{R^2},\qquad
	\tilde{Z}=\frac{(r-M)x}{\pa{1-x^2}\Delta+(r-M)^2x^2}
	=\frac{Z}{\Pi^2+Z^2}
	=\frac{Z}{R^2},
\end{align}
in terms of which they take the cleaner and $R$-separable---but not simply separable, due to an overall term---form
\begin{align}
	\label{eq:KerrSCT}
	\psi_-(\tilde{\Pi},\tilde{Z})=\sqrt{\tilde{\Pi}^2+\tilde{Z}^2}J_0(\lambda\tilde{\Pi})e^{-\lambda\tilde{Z}}.
\end{align}
The two families $\psi_+$ and $\psi_-$ are almost related by an interchange $(\Pi,Z)\leftrightarrow(\tilde{\Pi},\tilde{Z})$, but not exactly due to the $R$-factor $\sqrt{\tilde{\Pi}^2+\tilde{Z}^2}$.
Instead, they are related by a discrete inversion symmetry $S$ of Eq.~\eqref{eq:StataxiLaplacian} described in Sec.~\eqref{sec:Origin} below.

\subsection{Spin-\texorpdfstring{$s$}{s} perturbations}
\label{subsec:HigherSpinModes}

Stationary and axisymmetric Kerr perturbations of generic spin obey Eq.~\eqref{eq:StataxiTME}.
Its standard separable solutions are
\begin{align}
	\label{eq:UsualModes}
	\psi_\ell^{(s)}(r,x)=\Delta^{-\frac{1}{2}s}\br{a_\ell^{(s)}P_\ell^{(s)}\pa{\frac{r-M}{\sqrt{M^2-a^2}}}+b_\ell^{(s)}Q_\ell^{(s)}\pa{\frac{r-M}{\sqrt{M^2-a^2}}}}P_\ell^{(s)}(x),
\end{align}
where the $P_\ell^{(s)}(u)$ and $Q_\ell^{(s)}(u)$ are now associated Legendre polynomials and functions of the second kind, respectively.
Regularity at the poles now requires the integer $\ell\ge|s|$ to be at least as large as the spin.
As in the scalar case, $\ell$ is a multipole number, and the $P_\ell^{(s)}$ modes are regular at the horizon but not infinity, and vice versa for the $Q_\ell^{(s)}$ modes.
The existence of these separable solutions does not have a straightforward symmetry-based explanation \cite{Frolov2017}.

The spin-$s$ eigenmodes corresponding to the generators $H_i$ are solutions of Eq.~\eqref{eq:StataxiTME} that also satisfy the conditions%
\begin{subequations}
	\label{eq:Eigenmodes}
	\begin{align}
		H_0\psi_h^{(s)}(r,x)&=\pa{h+\frac{1}{2}}\psi_h^{(s)}(r,x),\\
		H_+\psi_+^{(s)}(r,x)&=-\lambda\psi_+^{(s)}(r,x),\\
		H_-\psi_-^{(s)}(r,x)&=+\lambda\psi_-^{(s)}(r,x).
	\end{align}
\end{subequations}
These infinite families of modes are explicitly expressed in terms of the Bessel function of the first kind $J_s(z)$ as
\begin{subequations}
	\label{eq:GeneralSpinModes}
	\begin{align}
		\psi_h^{(s)}(r,x)&=\Delta^{-\frac{1}{2}s}\br{\sqrt{\pa{1-x^2}\Delta+(r-M)^2x^2}}^hP_h^{(s)}\pa{\frac{(r-M)x}{\sqrt{\pa{1-x^2}\Delta+(r-M)^2x^2}}},\\
		\psi_+^{(s)}(r,x)&=\Delta^{-\frac{1}{2}s}J_s\pa{\lambda\sqrt{\pa{1-x^2}\Delta}}e^{-\lambda(r-M)x},\\
		\psi_-^{(s)}(r,x)&=\frac{\Delta^{-\frac{1}{2}s}}{\sqrt{\pa{1-x^2}\Delta+(r-M)^2x^2}}J_s\pa{\frac{\lambda\sqrt{\pa{1-x^2}\Delta}}{{\pa{1-x^2}\Delta+(r-M)^2x^2}}}e^{-\frac{\lambda(r-M)x}{\pa{1-x^2}\Delta+(r-M)^2x^2}}.
	\end{align}
\end{subequations}
Replacing $J_s(z)$ with the Bessel function of the second kind $Y_s(z)$ still provides solutions, but these blow up at $z=0$; thus, they are always irregular on the horizon and singular at the poles, which presumably renders them unphysical.

\section{Origin of the symmetry}
\label{sec:Origin}

The third main point of this paper is to explain the origin of the emergent $\mathsf{SL}(2,\mathbb{R})$ symmetry of Kerr perturbations.
The conformal symmetry generators \eqref{eq:Generators} of Eq.~\eqref{eq:StataxiTME} were originally discovered via a direct (and very tedious) application of Lie's algorithm for ``reduction of order'' using the \textsc{Mathematica} package \texttt{YaLie.m}.
In retrospect, however, a simpler approach to deriving the symmetries of the equation is to make use of the coordinate systems found in Sec.~\ref{sec:Separability} above.

The key idea is the following.
Scalar, stationary and axisymmetric Kerr perturbations are governed by Eq.~\eqref{eq:StataxiLaplacian}.
When $a=M=0$, the black hole vanishes altogether, leaving the simpler flat-space equation
\begin{align}
	\label{eq:FlatSpace}
	\pd_r\br{r^2\pd_r\psi(r,x)}+\pd_x\br{\pa{1-x^2}\pd_x\psi(r,x)}=0,
\end{align}
which, in light of Eq.~\eqref{eq:KerrLaplacian}, is manifestly the Laplace equation $\nabla^2\psi(r,x)=0$ for an axisymmetric massless scalar in Euclidean flat space $\mathbb{R}^3$ with spherical coordinates $(r,x=\cos{\theta},\phi)$.

Miraculously, when the full Kerr equation \eqref{eq:StataxiLaplacian} is expressed in the coordinates $(R,X)$, \textit{it takes the exact same form}:
\begin{align}
	\pd_R\br{R^2\pd_R\psi(R,X)}+\pd_X\br{\pa{1-X^2}\pd_X\psi(R,X)}=0.
\end{align}
Therefore, it immediately follows that any solution $\psi(r,x)$ of the flat-space equation \eqref{eq:FlatSpace} can be promoted to a solution of the full Kerr equation \eqref{eq:StataxiLaplacian} via the simple replacement $(r,x)\to(R,X)$ defined in Eq.~\eqref{eq:SphericalKerr}.
Said differently, one can exactly map the Kerr equation \eqref{eq:StataxiLaplacian} to its flat-space limit \eqref{eq:FlatSpace} via a coordinate transformation that effectively ``removes'' the black hole.
Thus, for stationary, axisymmetric perturbations, the Kerr spacetime appears to be ``flat''!

\subsection{Mapping to cylindrical coordinates}

In $\mathbb{R}^3$ with cylindrical coordinates $(\rho,z,\phi)$, the massless wave equation $\nabla^2\psi(\rho,z)=0$ for an axisymmetric scalar is
\begin{align}
	\label{eq:FlatLaplacian}
	\pa{\pd_\rho^2+\pd_z^2+\frac{1}{\rho}\pd_\rho}\psi(\rho,z)=0.
\end{align}
This equation is equivalent to Eq.~\eqref{eq:FlatSpace} under the transformation from cylindrical coordinates $(\rho,z)=\pa{r\sqrt{1-x^2},rx}$ to spherical coordinates $(r,x)=\pa{\sqrt{\rho^2+z^2},\frac{z}{\sqrt{\rho^2+z^2}}}$.
It can also be mapped into the Kerr equation \eqref{eq:StataxiLaplacian} by promoting
\begin{align}
	\label{eq:Mapping}
	(\rho,z)\to(\Pi,Z)=\pa{\sqrt{\pa{1-x^2}\Delta},(r-M)x},
\end{align}
consistent with Eq.~\eqref{eq:CartesianKerr} and the observation that $(\Pi,Z)=\pa{R\sqrt{1-X^2},RX}$ and $(R,X)=\pa{\sqrt{\Pi^2+Z^2},\frac{Z}{\sqrt{\Pi^2+Z^2}}}$.
The upshot is that all the properties of the Kerr equation \eqref{eq:StataxiLaplacian}, including its symmetries and solutions, can be derived in the context of the much simpler flat-space Laplace equation \eqref{eq:FlatLaplacian} and then mapped back to Kerr via the map \eqref{eq:Mapping}.

\subsection{Symmetries of the flat-space equation}
\label{subsec:Symmetries}

The flat-space equation \eqref{eq:FlatLaplacian} is a linear, second-order partial differential equation.
It is much easier to analyze than its mathematically equivalent (but seemingly more complex) Kerr form \eqref{eq:StataxiLaplacian}.
A straightforward application of Lie's algorithm to Eq.~\eqref{eq:FlatLaplacian} reveals that it admits precisely three continuous symmetries (in addition to the one related to the linear superposition principle, which states that $a\psi_1+b\psi_2$ is a solution whenever $\psi_1$ and $\psi_2$ are solutions).
Explicitly, given a physical solution $\psi(\rho,z)$, three new solutions may be obtained by acting with the differential operators
\begin{align}
	\label{eq:FlatSymmetries}
	H_+=\pd_z,\qquad
	H_0=\rho\pd_\rho+z\pd_z+\frac{1}{2},\qquad
	H_-=2\rho z\pd_\rho-\pa{\rho^2-z^2}\pd_z+z,
\end{align}
which generate the Lie algebra \eqref{eq:SL2R} of the global conformal group $\mathsf{SL}(2,\mathbb{R})$ with Casimir \eqref{eq:CasimirSL2R}.
In terms of $\mathcal{C}$, Eq.~\eqref{eq:FlatLaplacian} is
\begin{align}
	\label{eq:FlatCasimir}
	\pa{\mathcal{C}-\frac{1}{4}}\psi=0,
\end{align}
consistent with Eq.~\eqref{eq:Casimir} in the scalar case $s=0$.
Moreover, under the coordinate transformation \eqref{eq:Mapping}, the generators \eqref{eq:FlatSymmetries} map onto those in Eq.~\eqref{eq:Generators} with $s=0$.
This new form of the generators justifies their names in Sec.~\ref{sec:Separability}:
\begin{itemize}
	\item $H_+=\pd_z$ is a \textit{bona fide} vector field (and a Killing vector on $\mathbb{R}^3$) generating vertical translations along the $z$-axis;
	\item $H_0=D+\frac{1}{2}$ is closely related to the vector field $D=\rho\pd_\rho+z\pd_z$ generating flat-space dilatations; and finally,
	\item $H_-=K_z+z$ is closely related to the vector field $K_z=2\rho z\pd_\rho-\pa{\rho^2-z^2}\pd_z$ generating the special conformal transformation associated with translations in the $z$-direction (it is related to it via the inversion \eqref{eq:Inversion} below).
\end{itemize}

A few comments are now in order.
First, $\mathbb{R}^3$ admits exactly ten conformal Killing vectors (the maximal number in three dimensions): six exact ones and four conformal ones.
The exact ones consist of the spatial translations $P_x=\pd_x$, $P_y=\pd_y$, and $P_z=\pd_z$, together with the usual spatial rotations with generators $J_x$, $J_y$, and $J_z$ given above Eq.~\eqref{eq:Reducibility}.
The conformal Killing vectors are the dilatation $D$ and the special conformal transformations $K_x$, $K_y$, and $K_z$.
Of these ten conformal symmetries, only three preserve axisymmetry: $P_z$, $D$, and $K_z$, which are unsurprisingly the ones in the above list.
The other vectors are either trivial or else fail to preserve axisymmetry by introducing $\phi$-dependence.

Second, it is useful to note that the eigenstates $\phi_h(r,x)$ also diagonalize dilatations: they obey $D\phi_h(r,x)=h\phi_h(r,x)$.
Hence, the conformal weight $h$ is manifestly a scaling dimension.

Finally, a perhaps more intuitive way to understand the symmetries of Eq.~\eqref{eq:FlatLaplacian}, as well as the appearance of the $\frac{1}{2}$ shift in $H_0$, is to observe that under the joint field redefinition and coordinate transformation
\begin{align}
	\psi(\rho,z)=\frac{1}{\rho^p}\phi(\rho,z),\qquad
	p=\frac{1}{2},\qquad
	(\rho,z)\to(w,\bar{w})=(\rho+iz,\rho-iz),
\end{align}
the three-dimensional, axisymmetric Laplace equation \eqref{eq:FlatLaplacian} transforms into the two-dimensional wave equation
\begin{align}
	\label{eq:Holomorphic}
	\pd_w\pd_{\bar{w}}\phi(w,\bar{w})=-\frac{1}{4\pa{w+\bar{w}}^2}\phi(w,\bar{w})
\end{align}
for a massive field with position-dependent mass $m=\frac{1}{2\rho}$.
Without this mass term, this would be the two-dimensional Laplace equation, which admits infinitely many symmetries $f(w)\pd_w+\bar{f}(\bar{w})\pd_{\bar{w}}$ generating the local conformal group: setting $f(w)=-w^{n+1}$ gives generators $L_n=-w^{n+1}\pd_w-\bar{w}^{n+1}\pd_{\bar{w}}$ obeying the Witt algebra $[L_m,L_n]=(m-n)L_{m+n}$.
Roughly, the mass term breaks the infinite-dimensional local conformal symmetry down to a global subgroup $\mathsf{SL}(2,\mathbb{R})$.

\subsection{Finite symmetry transformations}

The finite action of $H_+$ is simply a vertical translation by a finite amount along the $z$-axis:
\begin{align}
	e^{\lambda H_+}\psi(\rho,z)=\psi(\rho,z+\lambda).
\end{align}
It is striking that such a symmetry persists even in the presence of a black hole, whose position manifestly breaks the naive version of translation invariance along its axis of rotation.
The finite action of $D=H_0-\frac{1}{2}$ is just the rescaling:
\begin{align}
	e^{\lambda D}\psi(\rho,z)=\psi\pa{e^{\lambda}\rho,e^{\lambda}z}.
\end{align}
The finite action of $H_-$ is more complicated (but as a check, note that $e^{\lambda H_-}e^{\lambda H_-}\psi(\rho,z)=e^{2\lambda H_-}\psi(\rho,z)$ as expected):
\begin{align}
	e^{\lambda H_-}\psi(\rho,z)=\frac{1}{\sqrt{\lambda^2\rho^2+\pa{1-\lambda z}^2}}\psi\pa{\frac{\rho}{\lambda^2\rho^2+\pa{1-\lambda z}^2},\frac{z-\lambda\pa{\rho^2+z^2}}{\lambda^2\rho^2+\pa{1-\lambda z}^2}}.
\end{align}
There is also a discrete inversion symmetry, which takes
\begin{align}
	\label{eq:DiscreteInversion}
	\psi(\rho,z)\longrightarrow S\psi(\rho,z)
	=\frac{1}{\sqrt{\rho^2+z^2}}\psi\pa{\frac{\rho}{\rho^2+z^2},\frac{z}{\rho^2+z^2}}.
\end{align}
This transformation is an involution, meaning that it is self-inverse: it squares to the identity, $S^2=1$.
One can check that the special conformal transformation is a composition of inversion, translation, and inversion:
\begin{align}
	\label{eq:Inversion}
	e^{\lambda H_-}=Se^{-\lambda H_+}S.
\end{align}
This inversion is even simpler in spherical coordinates, where it is manifestly an inversion through the sphere $r=1$:
\begin{align}
	\label{eq:RadialInversion}
	S\psi(r,x)=\frac{1}{r}\psi\pa{\frac{1}{r},x}.
\end{align}

\subsection{Separable solutions}

As in Sec.~\ref{sec:Separability}, each of the generators of $\mathsf{SL}(2,\mathbb{R})$ is associated with a coordinate system in which Eq.~\eqref{eq:FlatLaplacian} separates.
For $H_0$, these are the spherical coordinates $(r,x)$, in which the separable solutions are
\begin{align}
	\label{eq:FlatSpaceModes}
	\psi_h(r,x)=r^hP_h(x).
\end{align}
For $H_+$, these are the cylindrical coordinates $(\rho,z)$, in which the separable solutions are
\begin{align}
	\label{eq:SimplePlus}
	\psi_+(\rho,z)=J_0(\lambda\rho)e^{-\lambda z}.
\end{align}
For $H_-$, these are the inverse cylindrical coordinates $(\tilde{\rho},\tilde{z})=\pa{\frac{\rho}{r^2},\frac{z}{r^2}}$, which give rise to the $R$-separable solutions
\begin{align}
	\label{eq:SimpleMinus}
	\psi_-(\rho,z)=\frac{1}{\sqrt{\rho^2+z^2}}J_0\pa{\frac{\lambda\rho}{\rho^2+z^2}}e^{-\frac{\lambda z}{\rho^2+z^2}}.
\end{align}
Each of these separable solutions diagonalizes its associated generator:
\begin{align}
	H_+\psi_+=-\lambda\psi_+,\qquad
	H_-\psi_-=\lambda\psi_-,\qquad
	H_0\psi_h=\pa{h+\frac{1}{2}}\psi_h.
\end{align}
Promoting $(r,x)\to(R,X)$ [or equivalently, $(\rho,z)\to(\Pi,Z)$] maps these solutions to their Kerr analogues in Sec.~\ref{sec:Separability}.

As noted below Eq.~\eqref{eq:KerrSCT}, the modes $\psi_+$ and $\psi_-$ are related by a discrete symmetry via the inversion \eqref{eq:DiscreteInversion}:
\begin{align}
	\label{eq:Interchange}
	S\psi_\pm(\rho,z)=\psi_\mp(\rho,z).
\end{align}
Additional interesting properties and special solutions of Eq.~\eqref{eq:FlatLaplacian}---and hence of Eq.~\eqref{eq:StataxiLaplacian}---are relegated to App.~\ref{app:AdditionalProperties}.

\section{Kerr tidal Love numbers vanish by conformal symmetry}
\label{sec:NoLove}

The fourth and final main point of this paper is to explain why a black hole is not tidally deformable---or equivalently, ``why its Love numbers vanish''---from a symmetry perspective.
The key takeaway here is that the $\mathsf{SL}(2,\mathbb{R})$ symmetry of Kerr perturbations identified in Sec.~\ref{sec:Symmetry} implies that black holes do not respond to axisymmetric tidal perturbations.
The basic argument is that applied tidal perturbations and their induced tidal responses live in separate irreducible (highest-weight) representations of this $\mathsf{SL}(2,\mathbb{R})$, and as a result are barred from mixing by this conformal symmetry.

\subsection{Tidal perturbations and their Love numbers}

Before presenting this concise argument, it is helpful to briefly review tidal perturbations and their Love numbers.
Many excellent discussions of this topic exist in the literature; see, for example, \cite{LeTiec2021} for a particularly clear treatment.

Consider an isolated, compact object that is embedded in an external gravitational field.
In Newtonian theory, the gravitational potential associated with this configuration is static (in equilibrium) and admits the mode decomposition
\begin{align}
	\label{eq:NewtonianPotential}
	U(r,x)=\sum_{\ell=0}^\infty\br{C_\ell r^\ell+\frac{\Lambda_\ell}{r^{\ell+1}}}P_\ell(x).
\end{align}
Here, the modes are also assumed to be axisymmetric for simplicity.
In this setting, the term scaling like $r^\ell$ corresponds to an applied $\ell$-pole tidal field, while the term scaling like $r^{-(\ell+1)}$ corresponds to the induced $\ell$-pole tidal response of the body.
In other words, the coefficients $C_\ell$ control the external tidal perturbations of the compact object, while the coefficients $\Lambda_\ell$ govern its internal tidal response.
Their ratios determine the tidal deformability of the object, that is, its propensity to respond to such tidal perturbations.
These ratios give rise to so-called ``Love numbers'' whose precise definition is subject to some ambiguity \cite{Gralla2018} and depends on the exact setting \cite{LeTiec2021}.
This issue is of no concern here: an object for which the coefficients $\Lambda_\ell$ vanish has no tidal response, or equivalently, it has vanishing Love numbers.

In the general-relativistic context, tidal perturbations of a rotating black hole are captured by the static spin-$s$ Kerr perturbations.
In axisymmetry, these are governed by Eq.~\eqref{eq:StataxiTME}.
As reviewed in Sec.~\ref{sec:Separability} [Eq.~\eqref{eq:UsualModes}], the usual modes are
\begin{align}
	\label{eq:BadModes}
	\psi_\ell^{(s)}(r,x)=\Delta^{-\frac{1}{2}s}\br{a_\ell^{(s)}P_\ell^{(s)}\pa{\frac{r-M}{\sqrt{M^2-a^2}}}+b_\ell^{(s)}Q_\ell^{(s)}\pa{\frac{r-M}{\sqrt{M^2-a^2}}}}P_\ell^{(s)}(x).
\end{align}
These modes are the Kerr analogue of the terms appearing in the Newtonian decomposition \eqref{eq:NewtonianPotential}.
The problem with these modes is that they no longer enforce a sharp separation between applied (external) perturbations and their induced (internal) response, as the positive and negative powers $r^\ell$ and $r^{-(\ell+1)}$ may mix in the $P_\ell^{(s)}(u)$ and $Q_\ell^{(s)}(u)$.

\subsection{Multipoles and highest-weight representations in flat space}

The root cause of this issue is that the standard separable modes \eqref{eq:BadModes} that appear in the usual decomposition for Kerr perturbations do not transform simply under dilations, unlike their flat-space analogues in the mode sum \eqref{eq:NewtonianPotential}.
More precisely, that sum can be rewritten in terms of the flat-space modes $\psi_h(r,x)$ defined in Eq.~\eqref{eq:FlatSpaceModes} as
\begin{align}
	\label{eq:ManifestSymmetry}
	U(r,x)=\sum_{h=0}^\infty\br{C_h\psi_h(r,x)+\Lambda_h\psi_{-(h+1)}(r,x)}.
\end{align}
This makes several important points clear.
First, in flat space, the $H_0$-eigenstates are precisely the usual multipoles: the positive-weight modes $\psi_\ell(r,x)$ are the external $\ell$-poles, while the negative-weight modes $\psi_{-(\ell+1)}(r,x)$ are the internal $\ell$-poles.
Second, each $\ell$-pole has a definite scaling dimension, and can therefore be uniquely identified by its conformal weight.
Third, the external and internal multipoles live in distinct irreducible representations of $\mathsf{SL}(2,\mathbb{R})$.

The lowest external multipole, the monopole $\psi_0(\rho,z)=1$, is both translation-invariant and dilatation-invariant.
In other words, it is annihilated by $H_+$, and the next higher external multipoles are its descendants with respect to $H_-$:
\begin{align}
	H_+\psi_0=0,\qquad
	H_0\psi_0=+\frac{1}{2}\psi_0,\qquad
	\psi_{h+1}\propto H_-\psi_h\quad(h\ge0).
\end{align}
Thus, all the external $\ell$-poles (with $h=\ell\ge0$) may be obtained by starting with the lowest-weight multipole $\psi_0$ (the monopole) and then increasing its weight with the ``raising'' operator $H_-$: the first descendant $\psi_1$ is the dipole, the next one is the quadrupole $\psi_2$, and so on.

Likewise, the ``lowest'' internal multipole, $\psi_{-1}(r,x)=\frac{1}{r}$, is invariant under special conformal transformations.
As such, it is annihilated by $H_-$, and the next higher internal multipoles are its descendants with respect to $H_+$:
\begin{align}
	H_-\psi_{-1}=0,\qquad
	H_0\psi_{-1}=-\frac{1}{2}\psi_{-1},\qquad
	\psi_{h-1}\propto H_+\psi_h\quad(h\le1).
\end{align}
Thus, all the internal multipoles (with $h<0$) can be obtained by starting with the highest-weight multipole $\psi_{-1}$ and then decreasing its weight with the ``lowering'' operator $H_+$.

In summary, the flat-space multipoles form two towers (irreducible representations) of $\mathsf{SL}(2,\mathbb{R})$-descendents: a lowest-weight representation with primary $\psi_0$, and a highest-weight representation with primary $\psi_{-1}$.
In common parlance, the distinction between ``highest'' and ``lowest'' weight is seldom drawn, and both types of representation are referred to as ``highest-weight'' (usually).
The raising and lowering operators $H_\pm$ change the weights of the modes (their scaling dimension), but not the Casimir $\mathcal{C}$ of the representations, which is always fixed at $\mathcal{C}=\frac{1}{4}$ by Eq.~\eqref{eq:FlatCasimir}.

As a final comment, observe that the discrete inversion \eqref{eq:RadialInversion} interchanges external and internal multipoles, since
\begin{align}
	S\psi_h(r,x)=\psi_{-(h+1)}(r,x),
\end{align}
due to the Legendre identity $P_{-(h+1)}(x)=P_h(x)$.
This is consistent with Eq.~\eqref{eq:Interchange} and the expansions of the solutions $\psi_+$ and $\psi_-$ used therein as sums over all the positive (or respectively, negative) multipoles given in App.~\ref{app:AdditionalProperties}.
However, $S$ is not continuously connected to the identity; such $\mathsf{SL}(2,\mathbb{R})$ transformations do not change the sign of the weight $h$.

\subsection{No Love in Kerr}

When generalizing from flat space to Kerr by replacing the modes $r^\ell$ and $r^{-(\ell+1)}$ with the usual modes in Eq.~\eqref{eq:BadModes}, the three desirable properties listed below Eq.~\eqref{eq:ManifestSymmetry} are all lost.
This is where the new symmetry uncovered in Sec.~\ref{sec:Symmetry} is crucial: to address this issue, one can instead use the basis of $H_0$-eigenstate solutions $\psi_h^{(s)}$ defined in Eqs.~\eqref{eq:EigenModesH0} and \eqref{eq:GeneralSpinModes}, which restores all these properties.
In particular, a spin-$s$ tidal perturbation of Kerr is then cast in the form
\begin{align}
	\psi^{(s)}(r,x)=\sum_{h=|s|}^\infty\br{C_h^{(s)}\psi_h^{(s)}(r,x)+\Lambda_h^{(s)}\psi_{-(h+1)}^{(s)}(r,x)},
\end{align}
where the tidal perturbations and their induced responses are once again clearly associated with different modes of opposite sign of the conformal weight, which live in distinct highest-weight irreps of $\mathsf{SL}(2,\mathbb{R})$.
The key point is worth reiterating here: the modes $\psi_h^{(s)}$ are labeled by their scaling dimension, $h$ which determines their $H_0$-eigenvalue [Eq.~\eqref{eq:Eigenmodes}], and the tidal sources and responses have different eigenvalues under $H_0$ (positive or negative $h$, respectively).

By the same reasoning as in flat space, it then follows that black holes have no tidal response (or vanishing Love numbers) in axisymmetry.
In other words, the fact (noted in Sec.~\ref{sec:Origin}) that the effects of a black hole can be negated by a coordinate transformation implies that its tidal response is identical to that of flat space, which has no tidal deformability since it is empty!

By contrast, the presence of a material body such as a star breaks conformal symmetry (indeed, a star cannot be removed by a change of coordinates), which is why this argument no longer holds and celestial bodies experience tides.

Mathematically, the main advantage of the eigenmodes $\psi_h^{(s)}$ over their predecessors \eqref{eq:BadModes} is that, rather than mixing modes $P_\ell^{(s)}$ and $Q_\ell^{(s)}$ with $\ell\ge|s|$ always positive, these new modes only use the Legendre polynomials $P_h^{(s)}$, but now with both signs of the conformal weight.
Nevertheless, they still provide a complete basis.
This is best illustrated by way of example.
Considering only the $(s,\ell)=(0,4)$ case for brevity, the modes \eqref{eq:EigenModesH0} with $h\ge0$ may be use to expand
\begin{align}
	P_4\pa{\frac{r-M}{\sqrt{M^2-a^2}}}P_4(x)=\sum_{h=0}^4\frac{c_h}{\br{\sqrt{M^2-a^2}}^h}\psi_h(r,x),\qquad
	(c_0,c_1,c_2,c_3,c_4)=\pa{\frac{3}{8},0,-\frac{15}{4},0,\frac{35}{8}}.
\end{align}
This amounts to a change of basis from the usual separable modes to the $H_0$-eigenmodes.
In general, the relation is much more complicated and involves an infinite (and possibly two-sided) sum over $h\ge0$ (and possibly $h<0$).

Curiously, in the flat-space limit $a,M\to0$, both sets of modes---the $H_0$-eigenstates $\psi_h$ and the usual separable solutions $\psi_\ell$---reduce to the same flat-space modes \eqref{eq:FlatSpaceModes}.
Thus, the distinction between them is only relevant for black holes, for which the usual modes can be extended in different ways.
That these different extensions degenerate in the flat-space limit is likely related to the disappearance of the ``hidden'' symmetry associated with the Carter tensor \eqref{eq:KillingTensor}. Or rather, this hidden symmetry reduces to the usual, geometrically realized, spherical symmetry, as in Eq.~\eqref{eq:Reducibility}.

\section{Discussion}

This section concludes the paper with a brief comparison to previous works and with some possible future directions.
First, it is interesting that different conformal symmetries have been found to emerge in various regimes of Kerr physics:
\begin{enumerate}
	\item in the low-energy (``soft'') limit $\omega M\to0$ for generic spin $a$,
	\item in the high-spin (``near-extremal'') limit $a\to M$ for perturbations near the superradiant bound $\omega\to m\Omega_H$,
	\item in the high-energy (``eikonal'') limit $\omega M\to\infty$ for generic spin $a$.
\end{enumerate}
The new $\mathsf{SL}(2,\mathbb{R})$ symmetry discovered herein falls under the first category.
It is associated with (axisymmetric) tidal perturbations, which have been the subject of much recent work: the large literature on this subject includes \cite{Charalambous2021a,Charalambous2021b,Hui2021,Hui2022,Berens2023}, with more references given in the introduction of \cite{Berens2023}.
An excellent summary of recent developments appears in \cite{Hui2022}.

The second $\mathsf{SL}(2,\mathbb{R})$ symmetry is associated with the emergence in the high-spin limit $a\to M$ of a ``throat-like'' region of divergent proper depth outside the horizon \cite{Bardeen1972}.
Global conformal symmetry is geometrically realized as the isometry group of this limiting Near-Horizon Extreme Kerr (NHEK) geometry \cite{Bardeen1999}, and its infinite-dimensional local extension is the asymptotic symmetry group preserving the boundary conditions at the ``mouth'' of the throat \cite{Guica2009}.
This observation is the basis for the conjectured holographic ``Kerr/CFT correspondence'' for (extreme) Kerr \cite{Guica2009,Bredberg2010}.

The third $\mathsf{SL}(2,\mathbb{R})$ symmetry acts on a specific class of waves: the quasinormal modes.
In the high-frequency limit, these waves can be approximated using null geodesics via the geometric-optics (or ``eikonal'') approximation \cite{Hadar2022,Fransen2023,Kapec2024}.
As a result, the symmetry lifts to the phase space of null geodesics, where it acts near the ``photon shell'' of bound photon orbits around the black hole---in particular, dilatations map successive images of the Kerr ``photon ring'' \cite{Hadar2022}.

An important challenge for the future is to elucidate the interrelations between these various emergent symmetries.
The work \cite{Castro2010} offers an attempt to do so by expressing a version of the conformal symmetry that emerges in the ``soft'' limit in a form similar to that of the one emerging in the ``near-horizon'' limit.
However, the low-energy realization of $\mathsf{SL}(2,\mathbb{R})$ identified in that paper is only approximate, and its generators are not physically well-defined due to their multi-valued dependence on the azimuthal coordinate $\phi$.
This prevents the symmetry from being solution-generating.
By contrast, the realization of $\mathsf{SL}(2,\mathbb{R})$ in Sec.~\ref{sec:Symmetry} assumes axisymmetry, but it is exact and maps solutions to solutions.

As a technical aside, it is interesting to note that the symmetry of \cite{Castro2010} was discovered using a type of ``conformal'' coordinates $(w_\pm,y)$ in terms of which the symmetry generators take the simple form of the $L_n$ defined below Eq.~\eqref{eq:Holomorphic}.
Likewise, the $\mathsf{SL}(2,\mathbb{R})$ generators \eqref{eq:Generators} were initially derived by applying Lie's algorithm directly to Eq.~\eqref{eq:StataxiTME}.
For this approach to be mathematically tractable, it was crucial to exchange the Boyer-Lindquist radius $r$ for a new coordinate
\begin{align}
	\tilde{r}=-2\pa{r-M+\sqrt{\Delta}},\qquad
	r=M-\frac{\tilde{r}}{4}-\frac{M^2-a^2}{\tilde{r}},
\end{align}
in terms of which the poloidal Kerr metric, with line element given by Eq.~\eqref{eq:Kerr} with $\ed t=\ed\phi=0$, is conformally flat:
\begin{align}
	ds_P^2=\Gamma\br{\ed\tilde{r}^2+\frac{\tilde{r}^2}{1-x^2}\ed x^2},\qquad
	\Gamma(\tilde{r},x)=\br{\pa{\frac{1}{2}-\frac{M}{\tilde{r}}}^2-\frac{a^2}{\tilde{r}^2}}^2+\frac{a^2x^2}{\tilde{r}^2}.
\end{align}
The application of Lie's algorithm simplifies in these coordinates, as the generators \eqref{eq:Generators} take a form similar to the $L_n$.

Finally, it is useful to compare the various forms of conformal symmetry that emerge in the ``soft'' limit (point 1).
The previous works \cite{Charalambous2021a,Charalambous2021b,Hui2021,Hui2022,Berens2023} tackled the static limit $\omega M\to0$ of the Teukolsky master equation \eqref{eq:TME}, without necessarily assuming axisymmetry.
In one version of the analysis \cite{Hui2021,Berens2023}, one removes the $t$-derivatives from Eq.~\eqref{eq:TME} but not its $\phi$-derivatives, resulting in a partial differential equation for fields depending on three coordinates $(r,x,\phi)$.
This PDE is then recast as the Laplace equation $\nabla^2\Phi(r,x,\phi)$ of an auxiliary three-dimensional spacetime.
The benefit of this approach is that the symmetries of this Laplacian correspond precisely to the ``melodic'' conformal Killing vector fields of this ``effective'' spacetime \cite{Berens2023}.
As discussed in Sec.~\ref{subsec:Symmetries}, there are never more than ten conformal Killing vector fields in three dimensions, so finding them is a relatively easier task, especially if the effective metric is simple.

By contrast, this approach fails when applied to the static and axisymmetric limit of the Teukolsky master equation \eqref{eq:TME} given in Eq.~\eqref{eq:StataxiTME}.
The reason is that there exists no ``effective'' two-dimensional metric for which this equation is the Laplacian (this is  straightforward to check directly).
Indeed, if this had been the case, then as discussed below Eq.~\eqref{eq:Holomorphic}, one may have expected Eq.~\eqref{eq:StataxiTME} to have an infinite-dimensional (local conformal) symmetry.

\textit{Note:} While this work was being finished, Penna reported on investigations along very similar lines \cite{Penna2025}.
While his paper follows a completely different
 approach, there is some overlap in results: the $\mathsf{SL}(2,\mathbb{R})$ that he identifies appears to correspond to the $s=0$ version of the one in Sec.~\ref{sec:Symmetry}, and he also realizes the significance of the mapping \eqref{eq:Mapping} used to explain the origin of this symmetry in Sec.~\ref{sec:Origin}.
In addition, Penna offers a complementary perspective that connects this $\mathsf{SL}(2,\mathbb{R})$ to the Geroch group.
On the other hand, the present paper derives the symmetry for generic spin $s\neq0$, solves for the associated bases of separable modes in Sec.~\ref{sec:Separability}, and connects their existence to Love numbers in Sec.~\ref{sec:NoLove}.

\acknowledgments

I am grateful to Roman Berens, Juan Maldacena, Leo Stein, and Andrew Strominger for their valuable comments.
In particular, I thank Juan Maldacena for suggesting the title of this paper, and for pointing out the connection between symmetry and Love numbers presented in Sec.~\ref{sec:NoLove}.
This paper was checked for correctness and logical consistency using \href{https://www.refine.ink}{refine.ink}.
My work is supported by the NSF grant AST-2307888 and by the NSF CAREER award PHY-2340457.

\appendix

\section{Additional properties of stationary axisymmetric Kerr perturbations}
\label{app:AdditionalProperties}

This appendix collects additional properties and solutions of Eq.~\eqref{eq:FlatLaplacian}, and by extension [via the map \eqref{eq:Mapping}], Eq.~\eqref{eq:StataxiLaplacian}.

\subsection{Generating functions}

The separable solutions $\psi_\pm$ defined in Eqs.~\eqref{eq:SimplePlus}--\eqref{eq:SimpleMinus} are the exponential generating functions for the internal and external multipoles \eqref{eq:FlatSpaceModes} that form two irreducible $\mathsf{SL}(2,\mathbb{R})$-representations (and also for the Legendre polynomials):
\begin{align}
	\label{eq:Series}
	\psi_+(\lambda,r,x)&=J_0\pa{\lambda r\sqrt{1-x^2}}e^{-\lambda rx}
	=\sum_{h=0}^\infty\frac{\pa{-\lambda}^h}{h!}\psi_h(r,x)
	=\sum_{n=0}^\infty\frac{\pa{-\lambda r}^n}{n!}P_n(x),\\
	\psi_-(\lambda,r,x)&=\frac{1}{r}J_0\pa{\frac{\lambda\sqrt{1-x^2}}{r}}e^{-\frac{\lambda}{r}x}
	=\sum_{h=0}^\infty\frac{\pa{-\lambda}^h}{h!}\psi_{-(h+1)}(r,x)
	=\frac{1}{r}\sum_{n=0}^\infty\frac{\pa{-\frac{\lambda}{r}}^n}{n!}P_n(x).
\end{align}
The regular generating function, which is of course also a solution of Eq.~\eqref{eq:FlatLaplacian}, is
\begin{align}
	\psi_\lambda(r,x)=\frac{1}{\sqrt{r^2+2\lambda rx+\lambda^2}} 
	=\frac{1}{r}\sum_{n=0}^\infty\pa{-\frac{\lambda}{r}}^nP_n(x)
	=\frac{1}{\lambda}\sum_{n=0}^\infty\pa{-\frac{r}{\lambda}}^nP_n(x).
\end{align}
As always, the exponential and regular generating functions are related by a Laplace transform: explicitly,
\begin{align}
	\psi_\lambda(r,x)=\int_0^\infty\psi_-(\lambda t,r,x)e^{-t}\ed t
	=\frac{1}{\lambda}\int_0^\infty\psi_+\pa{\frac{t}{\lambda},r,x}e^{-t}\ed t.
\end{align}

\subsection{Integral transformations}

It is possible to obtain all of the individual multipole solutions---namely, the $H_0$-eigenstates \eqref{eq:FlatSpaceModes}---from a single one of the separable solutions $\psi_\pm$.
Consider for instance how this works for $\psi_+(\rho,z)=J_0(\lambda\rho)e^{-\lambda z}$.
By Eq.~\eqref{eq:Series}, the coefficients of the series expansion of $\psi_+(\rho,z)$ in powers of $\lambda$ are all of the external multipoles $\psi_h(\rho,z)$ with $h\ge0$.
As for the internal ones, they may be recovered from an integral over $\lambda^n\psi_+(\rho,z)$ for an appropriate $n$.
For instance,
\begin{align}
	\psi_{-1}(\rho,z)=\frac{1}{\sqrt{\rho^2+z^2}}
	=\int_0^\infty\psi_+(\rho,z)\ed\lambda.
\end{align}
Likewise,
\begin{align}
	\psi_{-2}(\rho,z)=\frac{z}{\pa{\rho^2+z^2}^{3/2}}
	=\int_0^\infty\lambda\psi_+(\rho,z)\ed\lambda,
\end{align}
and more generally,
\begin{align}
	\psi_{-n}(\rho,z)\propto\int_0^\infty\lambda^{n-1}\psi_+(\rho,z)\ed\lambda.
\end{align}

\subsection{Green's function}

A Green's function for Eq.~\eqref{eq:FlatLaplacian} may be given in terms of the complete elliptic integral of the first kind $K(m)$ by
\begin{align}
	G\pa{\rho,z;\rho',z'}=\frac{1}{2\pi^2}\frac{1}{\sqrt{\pa{\rho-\rho'}^2+\pa{z-z'}^2}}K\pa{-\frac{4\rho\rho'}{\pa{\rho-\rho'}^2+\pa{z-z'}^2}}.
\end{align}
(Here, $m=k^2$ is the parameter, not the modulus $k$.)
The boundary conditions obeyed by this particular Green's function may be read off from its expansion as a sum over products of modes
\begin{align}
	G\pa{\rho,z;\rho',z'}=\frac{1}{4\pi}\sum_{h=0}^\infty\psi_h(\rho,z)\psi_{-(h+1)}(\rho',z'),
\end{align}
where one leg of the two-point function is always evaluated with internal multipoles and the other with external ones.

\subsection{General solutions}

It is possible to represent the general solution of Eq.~\eqref{eq:FlatLaplacian} in a very explicit form.
Since Eq.~\eqref{eq:FlatLaplacian} can be written as
\begin{align}
	\pd_z^2\psi=D_\rho\psi,\qquad
	D_\rho=-\pd_\rho^2-\frac{1}{\rho}\pd_\rho,
\end{align}
it follows that its general solution subject to the boundary conditions $\psi(\rho,0)=\Psi_0(\rho)$ and $\pd_z\psi(\rho,0)=\Psi_1(\rho)$ is thus
\begin{align}
	\label{eq:Form1}
	\psi(\rho,z)=\cosh\pa{z\sqrt{D_\rho}}\Psi_0(\rho)+\sinh\pa{z\sqrt{D_\rho}}\Psi_1(\rho).
\end{align}
An alternative way of rewriting the equation is in terms of the inverse cylindrical coordinates $(\tilde{\rho},\tilde{z})=\pa{\frac{\rho}{r^2},\frac{z}{r^2}}$:
\begin{align}
	\pd_{\tilde{z}}^2\tilde{\psi}=\tilde{D}_{\tilde{\rho}}\tilde{\psi},\qquad
	\tilde{\psi}=\sqrt{\rho^2+z^2}\psi,\qquad
	\tilde{D}_{\tilde{\rho}}=-\pd_{\tilde{\rho}}^2-\frac{1}{\tilde{\rho}}\pd_{\tilde{\rho}}.
\end{align}
Hence, another general solution subject to boundary conditions $\tilde{\psi}(\tilde{\rho},0)=\tilde{\Psi}_0(\tilde{\rho})$ and $\pd_{\tilde{z}}\tilde{\psi}(\tilde{\rho},0)=\tilde{\Psi}_1(\tilde{\rho})$ is thus
\begin{align}
	\label{eq:Form2}
	\tilde{\psi}(\tilde{\rho},\tilde{z})=\cosh\pa{\tilde{z}\sqrt{\tilde{D}_{\tilde{\rho}}}}\tilde{\Psi}_0(\tilde{\rho})+\sinh\pa{\tilde{z}\sqrt{\tilde{D}_{\tilde{\rho}}}}\tilde{\Psi}_1(\tilde{\rho}).
\end{align}
Going back to regular cylindrical coordinates, this is
\begin{align}
	\psi(\rho,z)=\frac{1}{\sqrt{\rho^2+z^2}}\br{\cosh\pa{\frac{z}{\rho^2+z^2}\sqrt{\tilde{D}_{\tilde{\rho}}}}\tilde{\Psi}_0\pa{\frac{\rho}{\rho^2+z^2}}+\sinh\pa{\frac{z}{\rho^2+z^2}\sqrt{\tilde{D}_{\tilde{\rho}}}}\tilde{\Psi}_1\pa{\frac{\rho}{\rho^2+z^2}}}.
\end{align}

The positive-weight multipoles $\psi_h(r,x)=r^hP_h(x)$ with $h\ge0$ are all generated from some seed $\Psi_0(\rho)$ with $\Psi_1(\rho)=0$ (if $h$ is even) or from a seed $\Psi_1(\rho)$ with $\Psi_0(\rho)=0$ (if $h$ is odd).
In either case, the seed is annihilated by $D_\rho^n$ raised to some positive even power $n$, so that the expansion in Eq.~\eqref{eq:Form1} terminates at finite order.
For example, $\psi_2(r,x)=r^2P_2(x)=z^2-\frac{1}{2}\rho^2$ is of the form \eqref{eq:Form1} with $\Psi_1(\rho)=0$ and seed $\Psi_0(\rho)=-\frac{1}{2}\rho^2$, which obeys $D_\rho^2\Psi_0=0$.

Likewise, the negative-weight multipoles $\psi_h(r,x)=r^hP_h(x)$ with $h<0$ are all generated from some seed $\tilde{\Psi}_0(\tilde{\rho})$ with $\tilde{\Psi}_1(\tilde{\rho})=0$ (if $h$ is even) or from a seed $\tilde{\Psi}_1(\tilde{\rho})$ with $\tilde{\Psi}_0(\tilde{\rho})=0$ (if $h$ is odd).
Either way, the seed is annihilated by $\tilde{D}_{\tilde{\rho}}^n$ raised to some positive even power $n$, so that the expansion in Eq.~\eqref{eq:Form2} terminates at finite order.
For example, $\psi_{-3}(r,x)=\frac{P_{-3}(x)}{r^3}=\sqrt{\tilde{\rho}^2+\tilde{z}^2}\pa{\tilde{z}^2-\frac{1}{2}\tilde{\rho}^2}$, so that $\tilde{\psi}_{-3}(\tilde{\rho},\tilde{z})=\tilde{z}^2-\frac{1}{2}\tilde{\rho}^2$ is of the form \eqref{eq:Form2} with $\tilde{\Psi}_1(\tilde{\rho})=0$ and seed $\tilde{\Psi}_0(\tilde{\rho})=-\frac{1}{2}\tilde{\rho}^2$, which obeys $\tilde{D}_{\tilde{\rho}}^2\tilde{\Psi}_0=0$.

\subsection{Spin-\texorpdfstring{$s$}{s} perturbations}

Finally, the flat-space generalization to generic spin $s$ of Eq.~\eqref{eq:FlatLaplacian} is simply Eq.~\eqref{eq:StataxiTME} with $a=M=0$.
Its symmetry generators may be read off from Eq.~\eqref{eq:Generators} with $a=M=0$.
Likewise, the separable solutions diagonalizing these generators are given by Eq.~\eqref{eq:GeneralSpinModes} with $a=M=0$.

\bibliographystyle{utphys}
\bibliography{KerrSL2R.bib}

\end{document}